\documentclass[12pt]{iopart}
\usepackage{amssymb}
\usepackage{cancel}
\usepackage{color,graphicx}

\usepackage{amsmath}
\usepackage{amsbsy}
\usepackage{amsthm}
\usepackage{bbm}
\usepackage{bm}
\usepackage{epsfig}
\usepackage{lscape}
\usepackage{float}
\usepackage{subfigure}
\usepackage{dcolumn}
\usepackage{color,epstopdf}
\usepackage{amscd}
\usepackage{amsfonts}  
\usepackage{mathrsfs}
\usepackage{verbatim}
\usepackage[]{cases}
\usepackage{wasysym}
\usepackage[utf8]{inputenc}
\usepackage[T1]{fontenc}
\usepackage{mathtools}

\newcommand\ro{\hat\rho}
\newcommand\Ho{\hat H}

\newcommand\po{\hat p}
\newcommand\zo{\hat z}

\newcommand\Lo{\hat L}

\newcommand\half{\frac{1}{2}}

\newcommand\beq{\begin{equation}}
\newcommand\eeq{\end{equation}}
\newcommand\beqa{\begin{eqnarray}}
\newcommand\eeqa{\end{eqnarray}}

\begin{document}
\title{Optimal control for feedback cooling in cavityless levitated optomechanics}
\author{Luca Ferialdi$^1$, Ashley Setter$^1$, Marko Toro\v{s}$^1$, Chris Timberlake$^1$ and Hendrik Ulbricht$^1$}
\address{$^1$ Department of Physics and Astronomy, University of Southampton, SO17 1BJ, United Kingdom}
\ead{ferialdi@ts.infn.it, A.Setter@soton.ac.uk, M.Toros@soton.ac.uk, ct10g12@soton.ac.uk and H.Ulbricht@soton.ac.uk}
\begin{abstract}
We consider feedback cooling in a cavityless levitated optomechanics setup, and we investigate the possibility to improve the feedback implementation. We apply optimal control theory to derive the optimal feedback signal both for quadratic (parametric) and linear (electric) feedback. We numerically compare optimal feedback against the typical feedback implementation used for experiments. In order to do so, we implement a tracking scheme that takes into account the modulation of the laser intensity. We show that such a tracking implementation allows us to increase the feedback strength, leading to faster cooling rates and lower center-of-mass temperatures.
\end{abstract}

\maketitle

\section{Introduction}

The ability to precisely control and cool the motion of mechanical resonators in order to generate quantum states is of great interest for testing fundamental physics, such as investigating the quantum-to-classical transition ~\cite{Basetal13, Aspetal14}. A wide variety of resonator systems have shown promise for achieving such goals, including membranes~\cite{Teufel11, Rosetal}, micro- and nano-resonators~\cite{Lucas17, Guo17, Malz18, Chan11} and cantilevers~\cite{Vinante17, Poggio07}. Although ground state cooling as been experimentally realized in optomechanical systems~\cite{Teufel11, Rosetal, Chan11}, there is an appetite to reach such states in levitated systems. Levitated nanoparticles are extremely well isolated from their environment, opening up the possibility for very long decoherence times and ground state cooling in room temperature conditions. Indeed, optically levitated silica particles have had their center-of-mass motion cooled to millikelvin~\cite{Vovetal, Gieseler2012, Setter2018, Li11} and sub-millikelvin~\cite{Vijay2016, Teb18} temperatures, whereas nanodiamonds~\cite{Rahetal, Franetal} have been used for spin coupling experiments~\cite{Neukirch15, Hoang16b}. Other levitation mechanisms, such as Paul traps~\cite{Alda16}, hybrid electro-optical traps~\cite{paul}, and magnetic traps~\cite{Houlton18, Obrien19, Slevak18} have also been proposed as candidates for preparing macroscopic quantum states~\cite{Goldetal, Walker19, Romero-Isart12} and testing spontaneous collapse models~\cite{Goldwater16, Vinante19}. In order for any of these resonator systems to approach the quantum regime, their motion must first be cooled to close to the ground state, which can be achieved with cryogenically cooling the environment or with active feedback schemes.  

\begin{figure}[t!]
  \centering
  \includegraphics[width=0.55\textwidth]{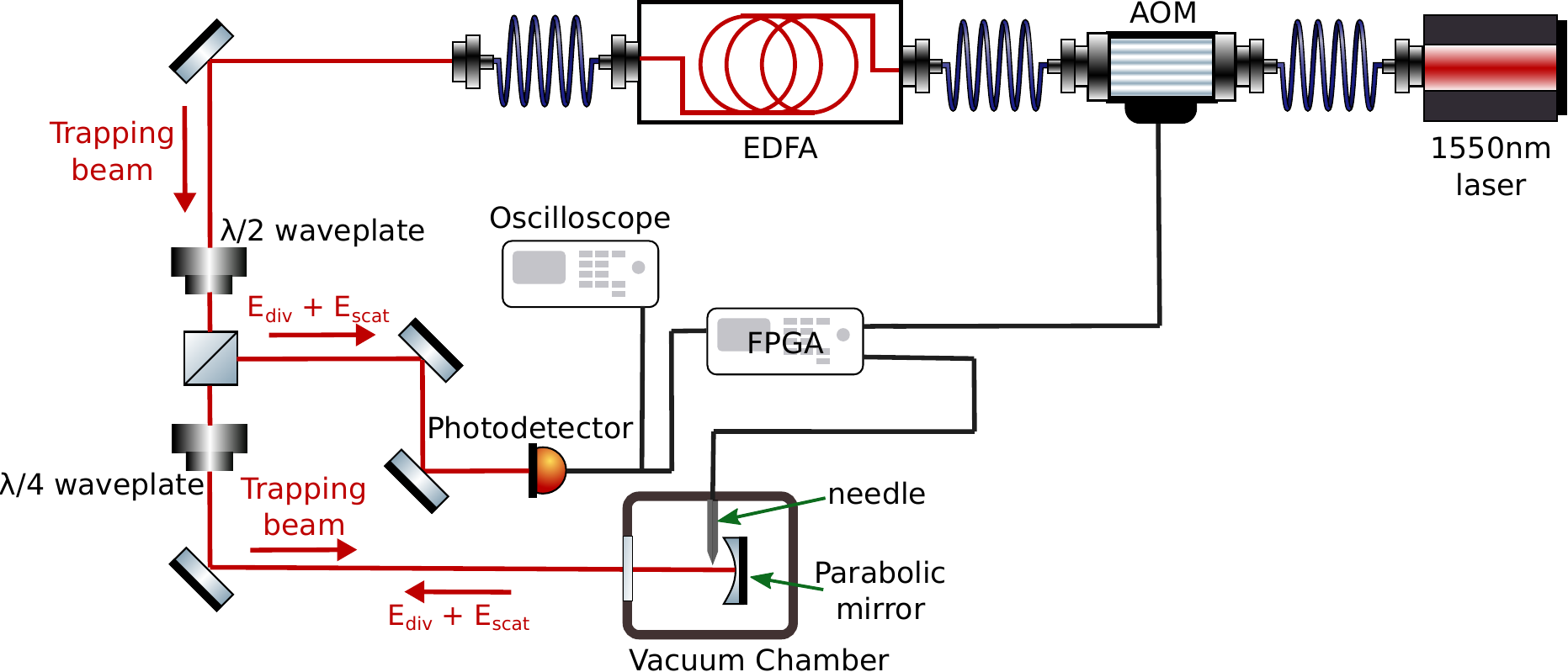}
  \caption{\label{setup} The experimental setup that we are simulating. The position of the particle is detected by interference between the scattered and divergent light at the photodetector. This detected signal is then passed into an oscilloscope for recording and a field programmable gate array (FPGA) to perform the tracking and generate the feedback signals which can then be sent to the acousto-optic modulator (AOM) to modulate the light and perform quadratic feedback or to the needle to perform linear feedback.}
  \end{figure}

In this paper we consider an optically levitated silica nanoparticle, trapped by by the gradient force generated by tightly focusing a 1550nm laser with a high numerical aperture (N.A.) paraboloidal mirror, as shown in figure~\ref{setup}. The optical trap is contained within a vacuum chamber to isolate the particle from its environment as much as possible. Typically, parametric (quadratic) feedback cooling, by modulating the intensity of the trapping laser at twice the particle's oscillation frequency~\cite{Vovetal, Gieseler2012}, is implemented to cool the particle's motion to $\sim$mK temperatures. Currently, feedback signals are implemented by tracking the phase of the oscillator by locking to the frequency of motion, using either lock-in amplifiers or, more recently, with a Kalman filter~\cite{Jaz70}. The Kalman filter, a sophisticated filtering technique used in engineering applications~\cite{kalman, Sandhu17, Wang17, Kara17}, can be implemented in real-time to accurately track the particle's motion, before applying the modulating feedback signal~\cite{Setter2018, Liao18}. Such schemes are very effective for tracking the particle motion for small laser modulation, but above a certain (low) threshold loses track of the particle. This is a limitation as higher modulation results in faster cooling rates and a lower final temperature.

Recently, cooling the motion of charged nanoparticles by applying an electric field which is at the same frequency of the particle's motion has been demonstrated~\cite{Teb18, Iwaetal} and implemented with optimal control protocols~\cite{Conangla19} for optical traps, as well as proposed for electrical traps~\cite{Goldetal}. A charged needle, placed in the vacuum chamber close to the laser focus, has been used for force sensing applications~\cite{Hemp17} and investigations of Fano resonances~\cite{Timberlake2019} in levitated optomechanics. To first approximation, the electric field generated by the needle couples linearly to the particle position, making it suitable to implement linear feedback cooling. By applying a force to oppose the particle motion, the amplitude of motion can be reduced. It is worth noting that for this cooling technique the coupling strength cannot be indefinitely high, as too strong an applied force would drive the particle to hotter temperatures, and could even result in the particle being ejected from the trap.   

In this article we consider whether it is possible to implement a feedback protocol which takes into account all the contributions to the particle dynamics, including decoherence and photon recoil, and compare to current feedback schemes discussed previously. We utilize optimal control theory to investigate both quadratic (parametric) and linear (electric) feedback (section~\ref{sec:Section:-feedback-schemes}). Optimal control theory has been applied to other experimental systems~\cite{Glaser15}, including for manipulation of Bose-Eintsein condensates to prepare complex quantum states~\cite{Mennemann15}, designing excitation pulses in NMR~\cite{Skinner03} and tailoring robustness in solid-state spin magnetometry~\cite{Nobauer15}. Additionally, it has been proposed for mixed state squeezing in cavity optomechanics~\cite{Basilwitsch19}, feedback cooling and squeezing of levitated nanopshperes in cavities~\cite{Genoni15} and recently for feedback cooling in low frequency magnetic traps~\cite{Walker19}. To compare optimal cooling with typical feedback cooling, we numerically emulate the system, by solving its equations of motion, and we track it by numerically solving a second set of equations (section~\ref{sec:Numerical analysis}). This tracking technique is more rubust, for our applications, than the one performed by a basic time-dependent Kalman filter, and allows us to increase the laser modulation to achieve better cooling of the trapped nanoparticle. Sections~\ref{subsec:Quad feedback} and \ref{subsec:Lin feedback} are respectively dedicated to quadratic and linear feedback results, while section~\ref{subsec:Discussion} concerns the common features of the two cooling schemes. In the next section we start by introducing the theoretical framework of the setup considered, and of its numerical simulation.

We remark that, although the analysis presented concerns a gradient force optical trap, the improved tracking developed is flexible, and could be also implemented in the other optomechanical setups previously described.

\section{Dynamical models\label{sec:Dynamical models}}
The optically levitated nanoparticle undergoes continuous monitoring of
its motion by the trapping laser~\cite{Hemp17,Toros18}. Specifically, we consider the experimental
situation when the translational and rotational degrees of freedom are
decoupled, and a single translation degree of freedom can be identified
in the detected signal~\cite{Wigner}. We will label the position of
this one-dimensional motion with $z$. We call $J$ the homodyne current that is physically accessible with the experimental setup, i.e. the quantity recorded by the measurement apparatus.

We consider two types of dynamical modelling of the experiment: (i) a tracking model
 and (ii) an emulation model~\cite{jaco}. In a nutshell, the models
of type (i)  are used to track the state of the system, and to provide its best estimate, either in real-time or in post-selection, given the \emph{input} homodyne current $J$ measured by the experimental setup~\cite{Doh00,Leon05}.
The only role of the models of type (ii) is to generate a trajectory of the system and to
\emph{output} the homodyne current $J$, i.e. to emulate the system when the experimentally measured $J$ is not available. This proves useful if one wants to investigate numerically the efficacy of a technique before implementing it in the experimental setup.

We now discuss in detail the tracking (section~\ref{subsec:System-tracking}) and emulation
(section~\ref{subsec:System-emulation}) model, while the feedback
mechanism will be discussed in section~\ref{sec:Section:-feedback-schemes}.

\subsection{System tracking\label{subsec:System-tracking}}
The dynamics of the continuously monitored trapped particle is described by the following master equation~\cite{Ralph18}:
\beq\label{me}
\frac{d\ro_T}{dt}=-\frac{i}{\hbar} [\Ho,\ro_T]+\left(2 k \mathcal{D}[\zo] +4\eta k (J-\langle\zo\rangle)\mathcal{H}[\zo]+\Gamma\mathcal{D}[\Lo]\right)\ro_T\,,
\eeq
where $\mathcal{D}[\Lo]\ro=\Lo\ro\Lo^\dag-\half\{\Lo^\dag\Lo,\ro\}$ and $\mathcal{H}[\zo]=\{\zo-\langle\zo\rangle,\ro\}$. The Hamiltonian $\Ho$ consists of two contributions $\Ho=\Ho_0+\Ho_{fb}$, with
\beq
H_0=\frac{\po^2}{2m}+\frac{m\omega^2}{2}\zo^2\,,\qquad H_{fb}=\beta u(t) \zo^2+ \delta v(t)\zo\,,
\eeq
the first being the harmonic trap provided by the laser, and the second being the feedback Hamiltonian, where $u(t)$ and $v(t)$ are feedback signals that depend on the particle measured position and  momentum. The second and third terms of equation~\eqref{me} describe respectively decoherence and monitoring provided by laser photons, with detection efficiency $\eta$ and monitoring strength  $k$. 
The fourth term accounts for decoherence due to residual gas particles, with $\Gamma=\frac{4mk_bT\gamma_c}{\hbar^2}$ and $\Lo=\zo+i\frac{\gamma_c}{\hbar\Gamma}\po$~\cite{cl,diss, Ferialdi17}. 
We will refer to this model as \emph{the tracking model} and to the corresponding state by $\rho_\text{\tiny{T}}$,
i.e. the tracked state. 

We assume that the particle is described by an initial Gaussian state. Since the dynamics of equation~\eqref{me} is quadratic, the system state remains Gaussian during its evolution, i.e. it is fully described by mean values of position and momentum operators ($\langle \zo\rangle_\text{\tiny{T}}$, $\langle \po\rangle_\text{\tiny{T}}$) and their variances ($V_z^\text{\tiny{T}}=\langle \zo^2\rangle_\text{\tiny{T}}-\langle \zo\rangle^2_\text{\tiny{T}}$, $V_p^\text{\tiny{T}}=\langle \po^2\rangle_\text{\tiny{T}}-\langle \po\rangle^2_\text{\tiny{T}}$, $C^\text{\tiny{T}}=\half\langle \{\zo,\po\}\rangle_\text{\tiny{T}}-\langle \zo\rangle_\text{\tiny{T}}\langle \po\rangle_\text{\tiny{T}}$). It is convenient to introduce the vector $\boldsymbol{x}\equiv(x_1,x_2,x_3,x_4,x_5)=(\langle \zo\rangle_\text{\tiny{T}},\langle \po\rangle_\text{\tiny{T}}, V_z^\text{\tiny{T}}, V_p^\text{\tiny{T}}, C^\text{\tiny{T}})$. By exploiting equation~\eqref{me} one finds that these evolve according to the following equations:
\beqa
\label{x1}\dot{x}_1&=&\frac{1}{m}\,x_2-\gamma_c\,x_1-8\eta\alpha(1+\beta u(t))\,(x_1-J)x_3\,,\\
\label{x2}\dot{x}_2&=&-m\omega^2(1+\beta u(t))\,x_1-\gamma_c\,x_2-8\eta\alpha(1+\beta u(t))\,(x_1-J)x_5\nonumber\\
&&+\delta v(t)\,,\\
\label{x3}\dot{x}_3&=&\frac{2}{m}\,x_5-2\gamma_c\,x_3+\frac{\gamma_c^2}{\Gamma}-8\eta\alpha(1+\beta u(t))\,x_3^2\,,\\
\dot{x}_4&=&-2m\omega^2(1+\beta u(t))\,x_5-2\gamma_c\,x_4+\hbar^2\Gamma-8\eta\alpha(1+\beta u(t))\,x_5^2\nonumber\\
&&+2\hbar^2\alpha(1+\beta u(t))\,,\\
\label{x5}\dot{x}_5&=&\frac{1}{m}\,x_4-m\omega^2(1+\beta u(t))\,x_3-2\gamma_c\,x_5-8\eta\alpha(1+\beta u(t))\,x_3x_5\,.
\eeqa
These equations account for the fact that the monitoring strength $k$ is proportional to the laser power, that is modulated by the feedback signal $u(t)$: $k=\alpha(1+\beta u(t))$,  where $\alpha=\frac{12\pi^2}{5\lambda^2}\frac{\sigma P}{\pi\omega_0^2\omega_L}$ is the coupling strength that depends on the laser power $P$~\cite{Setter2018}.

\subsection{System emulation\label{subsec:System-emulation}}

In the previous section we have discussed how to track the motion
of an optically levitated system. Specifically, we have assumed that
we are given an experimentally measured homodyne current $J$ which is then used to continuously
update our knowledge of the system. However, to \emph{emulate} the
system and to \emph{generate} an output homodyne current $J$ we have
to use a modified dynamical model. In particular, we will rewrite
the update term due to the detected photons as a stochastic back-action
term and we will include additional stochastic terms to account for
the undetected photons as well as gas collisions~\cite{Toros18, jaco, Rashid18}.
This is fully analogous to a classical emulation of the system: loosely
speaking, each scattering event due to photons (even if undetected)
or to gas particles makes the particle recoil, and this is modelled by
noise terms. We will refer to such a model as the \emph{emulation
model} and denote the corresponding state of the system by $\rho_\text{\tiny{E}}$,
i.e. the emulated state.

To emulate the system we consider the following dynamical equation:

\begin{alignat}{1}
d\hat{\rho}_\text{\tiny{E}}= & -\frac{i}{\hbar}[\hat{H},\hat{\rho}_\text{\tiny{E}}]\,dt+\Gamma\mathcal{D}[\hat{L}]\hat{\rho}_\text{\tiny{E}}\,dt+2k\mathcal{D}[\hat{z}]\hat{\rho}_\text{\tiny{E}}\,dt+\sqrt{2\eta k}\mathcal{H}[\hat{z}]\hat{\rho}_\text{\tiny{E}}\,dW\nonumber \\
 & +\sqrt{2(1-\eta)k}\mathcal{H}[\hat{z}]\hat{\rho}_\text{\tiny{E}}\,dV+\sqrt{\Gamma}\mathcal{H}[\hat{L}]\hat{\rho}_\text{\tiny{E}}\,dZ,\label{eq:emulation}
\end{alignat}
where the first line (from left to right) includes the unitary evolution, the diffusion terms due to gas scattering,
the diffusion term due to photon scattering, and the stochastic back-action
term due to the detected photons, where $dW$ is a Wiener process
with zero mean and correlation $\mathbb{E}[dWdW]=dt$. The third line
accounts for the nanoparticle recoil due to undetected photons ($\propto dV$)
and gas particles ($\propto dZ$), where $dV$ and $dZ$ are additional
independent Wiener processes with zero mean and correlations set to
$\mathbb{E}[dVdV]=\mathbb{E}[dZdZ]=dt$. This latter term $\propto dZ$
is responsible for thermalizing the motion of the nanoparticle with the gas particles. The associated homodyne current is given by:
\begin{equation}
Jdt=\langle\hat{z}\rangle_\text{\tiny{E}}dt+\frac{dW}{\sqrt{8\eta k}},\label{currentJ}
\end{equation}
where $\langle\,\cdot\,\rangle_\text{\tiny{E}}=\text{tr}[\,\cdot\,\hat{\rho}_\text{\tiny{E}}]$.

We now, similarly as for the tracking model, limit the discussion
to Gaussian states and introduce the vector $\boldsymbol{y}=(y_1,y_2,y_3,y_4,y_5)^T\equiv(\langle\zo\rangle_E,\langle\po\rangle_E,V^\text{\tiny{E}}_z,V^\text{\tiny{E}}_p,C^\text{\tiny{E}})$, where 
$V^\text{\tiny{E}}_z=\langle \zo^2\rangle_\text{\tiny{E}}-\langle \zo\rangle_\text{\tiny{E}}^2$, $V^\text{\tiny{E}}_p=\langle \po^2\rangle_\text{\tiny{E}}-\langle \po\rangle_\text{\tiny{E}}^2$, $C^\text{\tiny{E}}=\half\langle \{\zo,\po\}\rangle_\text{\tiny{E}}-\langle \zo\rangle_\text{\tiny{E}}\langle \po\rangle_\text{\tiny{E}}$). In this case equation~\eqref{eq:emulation} can be reduced
to the following coupled set of stochastic differential equations:
\beqa
\label{y1}\dot{y}_1&=&\frac{1}{m}\,y_2-\gamma_c\,y_1+\sqrt{8 \eta k}\,y_3\,dW\label{emx1}+\sqrt{8 (1-\eta) k}\,y_3\,dV\nonumber\\
&&+\sqrt{\Gamma}\left(2\,y_3-\frac{\gamma_{c}}{\Gamma}\right)\,dZ\,,\\
\label{y2}\dot{y}_2&=&-m\omega^2(1+\beta u(t))\,y_1-\gamma_c\,y_2+\delta v(t)+\sqrt{8 \eta k}\,y_5\,dW\nonumber\\
&&+\sqrt{8 (1-\eta)k}\,y_5\,dV+2\sqrt{\Gamma}\,y_5\,dZ\,,\\
\label{y3}\dot{y}_3&=&\frac{2}{m}\,y_5+2\gamma_c\,y_3-(8 k+4\Gamma)\,y_3^2\,,\\
\dot{y}_4&=&-2m\omega^2(1+\beta u(t))\,y_5-2\gamma_c\,y_4+\hbar^2(2k+\Gamma)-(8 k+4\Gamma)\,y_5^2\,,\\
\label{emx5}\dot{y}_5&=&\frac{1}{m}y_4-m\omega^{2}(1+\beta u(t))y_3-(8k+4\Gamma)y_3y_5\,.
\eeqa

\section{Optimal feedback\label{sec:Section:-feedback-schemes}}

Our aim is to determine the optimal controls $u^*(t)$ and $v^*(t)$ (here and in the following the asterisk denotes the optimal realization of a function) that provide the best cooling of the trapped particle, i.e. that minimize mean energy 
\beq\label{meanE}
\langle \Ho_0\rangle= \frac{1}{2m}\left(y_1^2+y_3\right)+\frac{m\omega^2}{2}\left(y^2_2+y_4\right)\,.
\eeq
Note that, although $\langle \Ho_0\rangle$ does not depend explicitly on the control functions $u(t)$ and $v(t)$, $\boldsymbol{y}$ does. Since such a dependence is linear, LQG optimization cannot be applied~\cite{DoJa99,wise}, and one needs to tackle the problem differently.
We exploit Pontryagin's Minimum Principle (PMP), an important tool of optimal control theory, that allows to find the optimal control that minimizes a given cost function~\cite{pontry}.  
The one solved by the PMP is a minimization problem with constraint (given by the equations of motion \eqref{x1}-\eqref{x5}). 
It is convenient to introduce the `co-states vector' $\boldsymbol{\lambda}\equiv(\lambda_1,\lambda_2,\lambda_3,\lambda_4,\lambda_5)$ and to define a `co-state Hamiltonian' as follows:
\beq
H_{co}(\boldsymbol{\lambda},\boldsymbol{y},u,v)=\boldsymbol{\lambda}\cdot\dot{\boldsymbol{y}}-\langle \Ho_0\rangle\,.
\eeq
One can check that the evolution equation for the co-states is
\beq\label{boldl}
\dot{\boldsymbol{\lambda}}=-\frac{\partial}{\partial \boldsymbol{y}} H_{co}(\boldsymbol{\lambda},\boldsymbol{y},u,v)\,,
\eeq
while equations~\eqref{x1}-\eqref{x5} can be conveniently rewritten as follows:
\beq\label{boldy}
\dot{\boldsymbol{y}}=\frac{\partial}{\partial \boldsymbol{\lambda}} H_{co}(\boldsymbol{\lambda},\boldsymbol{y},u,v)\,.
\eeq
Pontryagin's principle precisely states that the optimal control $u^*$, $v^*$ are those such that
\beq\label{cond}
H_{co}(\boldsymbol{\lambda^*},\boldsymbol{y^*},u^*,v^*)\leq H_{co}(\boldsymbol{\lambda^*},\boldsymbol{y^*},u,v)\,.
\eeq
Since the equations of motion for the components of $\boldsymbol{y}$ are linear both in $u$ and $v$, one can check the optimal signals satisfying the condition~\eqref{cond} are
\beq\label{controls}
u^*(t)=-\mathrm{sgn}\left[\boldsymbol{\lambda^*}\cdot\frac{\partial  H_{co}}{\partial \boldsymbol{u}}\right]\,,\qquad
v^*(t)=-\mathrm{sgn}\left[\boldsymbol{\lambda^*}\cdot\frac{\partial  H_{co}}{\partial \boldsymbol{v}}\right]\,,
\eeq
where $\mathrm{sgn}$ is the sign function that is $1\,(-1)$ when its argument in positive (negative).
We remark that the sign function form of the control has a simple intuitive explanation: in the case of quadratic feedback one wants to stiffen (weaken) the trap maximally when the particle moves away (towards) the trap center, and, similarly in the case of linear (electric) feedback one would like to stop the particle in its motion by applying the maximum ``breaking'' force. The only restriction is  thus on the
the feedback strength, i.e. on the modulation depth: if the  modulation depth is too strong one risks loosing track of the particle,or worse,  loosing the particle from the trap.
In order to obtain the explicit expressions for the two control functions, one needs to solve the two coupled sets of equations~\eqref{boldl}-\eqref{boldy}. This is in general a hard task because, while the state equations~\eqref{boldy} have \emph{initial} boundary conditions and propagate \emph{forward} in time, the co-state equations~\eqref{boldl} have \emph{final} boundary conditions (at the measurement time $\Delta t$) and propagate \emph{backward} in time~\cite{pontry}. As we will discuss in the next section, under certain conditions  it is numerically convenient to adopt a different strategy instead of solving the co-state equations.

\section{Numerical analysis of feedback schemes\label{sec:Numerical analysis}}
To emulate the system we first discretize equations~\eqref{currentJ}-\eqref{emx5}, i.e. we consider a time step $\Delta t_\text{\tiny{E}}$, and the Wiener increments $\Delta W$, $\Delta V$, $\Delta Z$. We set the initial state to be a Gaussian thermal state, i.e.
\beq
y_3 (t=0) = \frac{\hbar}{2 m \omega}\text{coth}\left(\frac{\hbar \omega}{2 k_B T}\right)\label{y3i}\,,\qquad y_4 (t=0) = \frac{\hbar m \omega}{2}\text{coth}\left(\frac{\hbar \omega}{2 k_B T}\right),
\eeq
and $y_1=y_2=y_5=0$, where $T$ is the temperature of the gas particles and $k_B$ is Boltzmann's constant. Specifically, we set $T=300$K, i.e. we assume that the gas of particles is at room temperature. We set $\omega=2\pi \times 70$kHz, and $m = 9.42\times10^{-19}$kg, which are typical values of trapped dielectric silica particles of radius $\sim50$nm. We also set $\eta = 0.003$ and $\alpha=4.04\times10^{25}$ which are typical values for our optical trap~\cite{Vovetal}. We then propagate the initial state for a time $t=t_\text{prep}\sim 5$ms, with the  control functions set to $u=v=0$: this preparation procedure ensures that the state of the system at time $t=t_\text{prep}$ is more realistic, i.e. the noise and dynamics of the emulation model will drive the system to a new state $\bm{y}(t_\text{prep})$. It also allows the tracking to converge and begin tracking the system well. Specifically, we solve the stochastic differential equations in equations~\eqref{currentJ}-\eqref{emx5} using the fourth-order stochastic Runge-Kutta method. 

\begin{figure}[t!]
  \centering
  \includegraphics[width=0.45\textwidth]{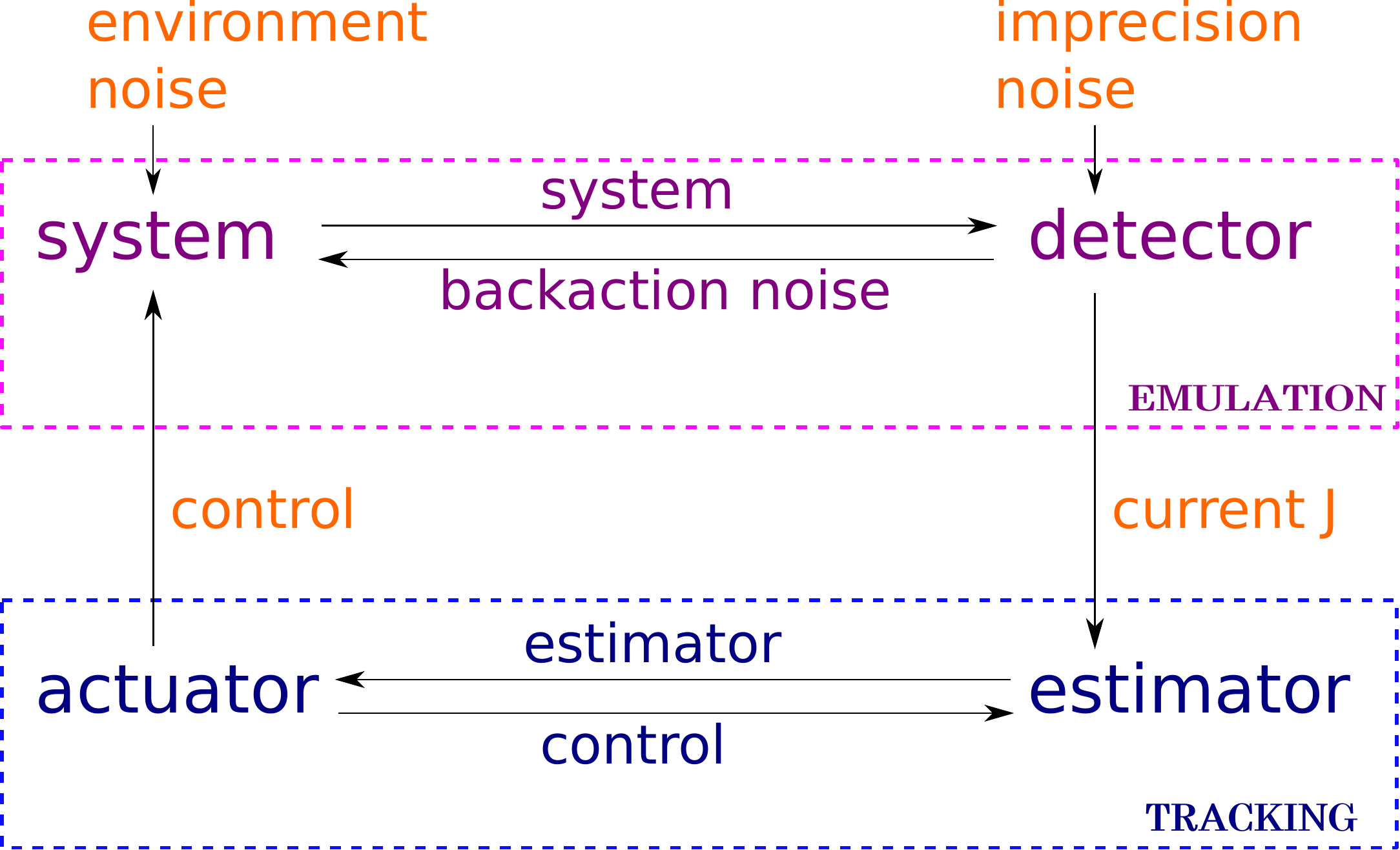}
  \caption{\label{figureS} Schematic diagram of the numerical simulation. The emulation part (purple) consists of the system and detector: the system is a Gaussian state described by the vector $\bm{y}$ and evolves according to equations~\eqref{emx1}-\eqref{emx5} with three inputs, i.e. the control ($u$ and $v$), the environmental noise ($dV$ and $dZ$), and the back-action noise ($dW$), while the detector has two inputs, i.e. the state $\bm{y}$ and the imprecision noise ($dW$), and produces the output current $J$ given by equation~\eqref{currentJ}. The estimation and control part (blue) consists of the tracking and actuator:  the estimator of the state consists of the vector $\bm{x}$ and evolves according to equations~\eqref{x1}-\eqref{x5} with two inputs, i.e. the control ($u$ and $v$) and the current $J$, while the actuator controls the functions $u$ and $v$ in response to the best estimate of the system given by $\bm{x}$. In an experimental realization the emulation part and simulated current $J$ are replaced by the experiment and the experimental current, respectively; the estimation and control part remains unchanged.}
\end{figure}

To track and control the system we need to solve in parallel also the equations~\eqref{x1}-\eqref{x5}, as well as choose the control functions $u$ and $v$. However the experimentally available time-step for the tracking and control is limited by the apparatus, e.g. sampling rates, reaction times and time lags. It is thus reasonable to consider larger time-steps, $\Delta t=M \Delta t_\text{\tiny{E}}$ and $\Delta t_\text{\tiny{C}}=\frac{\Delta t}{N}$, for the measurement and tracking/control respectively, with $N,M \in \mathcal{N}$. To simulate the current experimental capabilities presented in~\cite{Setter2018} we set $N=5$, $M=2000$ and $\Delta t_\text{\tiny{E}}=0.5$ns. We have verified numerically that such a value of $\Delta t_\text{\tiny{E}}$ provides with enough temporal resolution to simulate sufficiently well the evolution of the system. 

We set the initial state of the tracking at time $t=0$ to be a Gaussian thermal state, i.e. $\bm{x}(0)=\bm{y}(0)$, where the non-zero values of $\bm{y}(0)$ are given in equations~\eqref{y3i}. We switch on the feedback control at time $t=t_\text{prep}$. In order to avoid the difficulties of propagating the co-states equations backward in time, we adopt the following strategy. At each step $\Delta t$ we select the optimal control by selecting the optimal trajectory: we propagate the estimated state $\bm{x}$ forward in time for $\Delta t$ using the time-step $\Delta t_\text{\tiny{C}}=\frac{ \Delta t}{N}$ for each possible trajectory of the controls $u$ or $v$. Since according to equation~\eqref{controls} the value $u$ can have only values $\pm 1$ this amounts to $2^N$ trajectories; the same applies also for $v$. If both $u$ and $v$ would be controlled simultaneously in such a way we would have a total of $4^N$ trajectories. We select the trajectory that minimizes the cost function given in equation~\eqref{meanE}, i.e. the one that minimizes the estimated energy. It turns out that, at least for low values $N$, the  parallelization of the optimal control problem is computationally feasible. In particular, the scenario investigated here is particularly relevant for experiments involving FPGAs; for example, setting $N=5$ gives a total of $2^5=32$ trajectories for a single control function, which is readily solved in parallel using even moderately priced FPGAs. 

The schematic diagram in figure~\ref{figureS} gives an overview of the emulation-tracking implementation; for a more detailed introduction see e.g.~\cite{wise,jaco}. The feedback details for quadratic and linear cases will be respectively discussed in the following subsections~\ref{subsec:Quad feedback} and \ref{subsec:Lin feedback}. Section~\ref{subsec:Discussion} is devoted to the discussion of common features of the two feedback schemes.

\subsection{Quadratic feedback\label{subsec:Quad feedback}}
Parametric (i.e. quadratic) feedback is widely used in levitated optomechanics. The relevant equations describing this type of feedback can be obtained simply setting $\delta=0$ in section~\ref{sec:Dynamical models}. This type of feedback is typically performed by modulating the laser at twice the phase of the particle, setting $u= \frac{\omega}{E} x_1 x_2$~\cite{Setter2018}, where $E=\frac{x_2^2}{2m}+\frac{m\omega^2 }{2}x_1^2$. For a fixed value of $\beta$ we can then directly compare the optimal control $u^*$ with the simple double phase $u$. 
\begin{figure}[t!]
  \centering
  \includegraphics[width=0.45\textwidth]{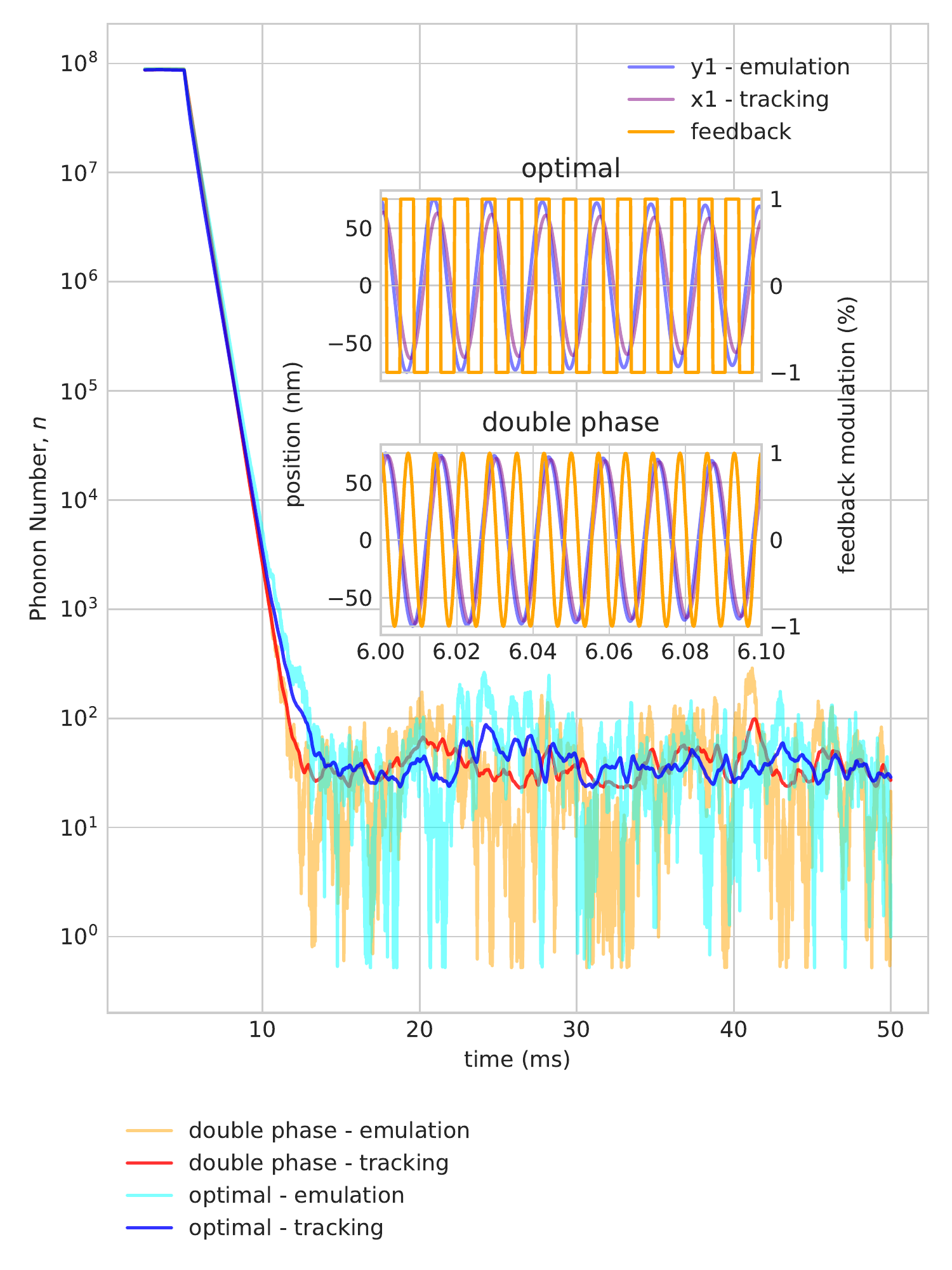}
  \caption{\label{figure1} Quadratic feedback: simulated time trace of the phonon number associated with the translational motion in the emulation and tracking equations. For the first $5\,ms$ the system evolves freely in the harmonic trap before optimal or double phase quadratic feedback cooling is applied. The insets show a small slice of the simulated position in the emulation and tracking equations along with the corresponding feedback signal applied to cool the translational motion for optimal and double phase quadratic feedback cooling. For these two simulations $\beta = 0.01$. The initial cooling rate $r_c$ is $\sim 2089s^{-1}$ where the phonon number $n = Ae^{(-r_c t)}$.
}
\end{figure}
\begin{figure}[t!]
  \centering
  \includegraphics[width=0.45\textwidth]{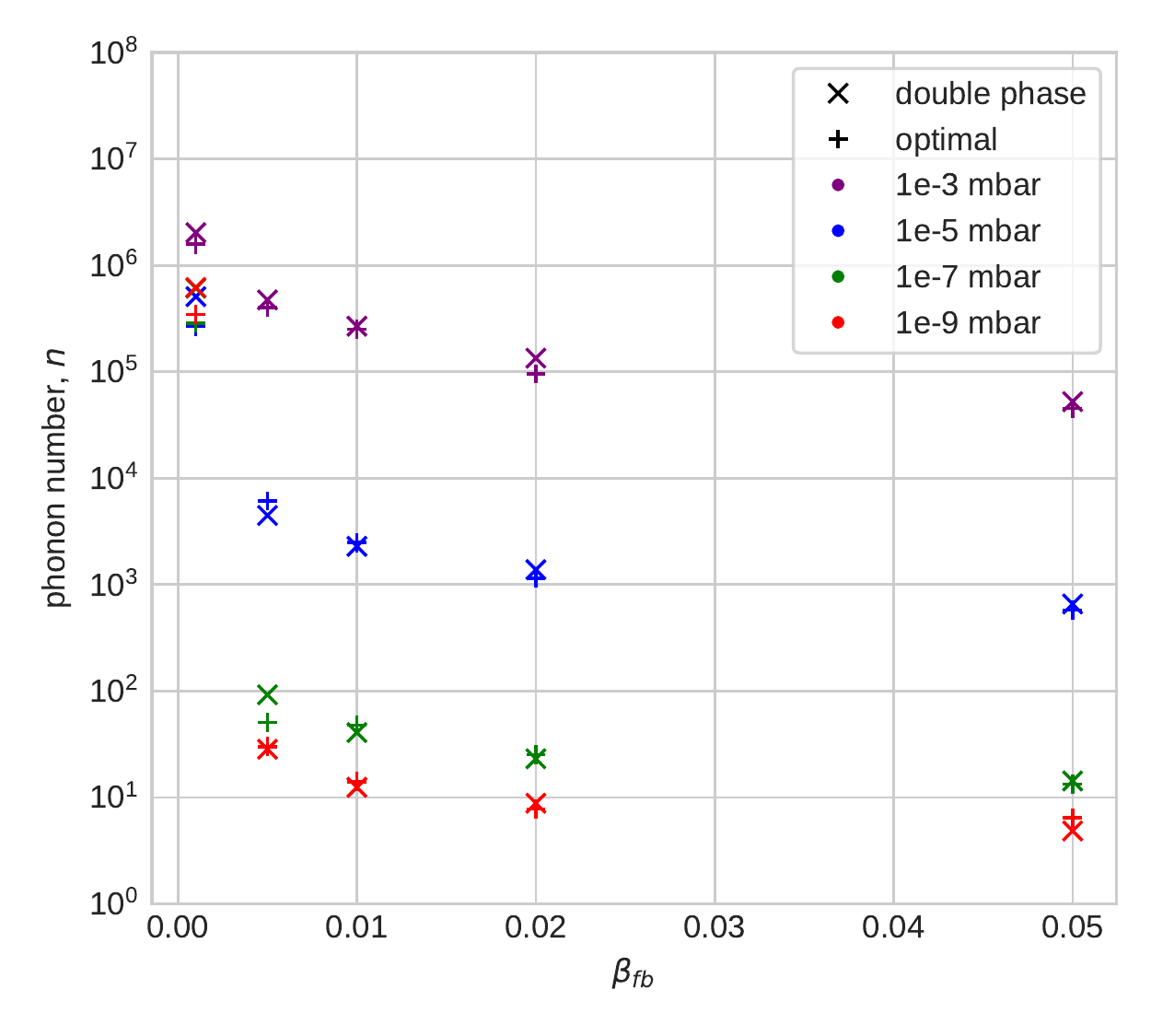}
  \caption{\label{figure2}
Quadratic feedback: shows the dependency of the average phonon number reached once the energy has converged with the modulation depth of feedback cooling for optimal and double phase quadratic feedback at different pressures.}
\end{figure}
Specifically, figures~\ref{figure1} and~\ref{figure2} show that the cooling obtained with the double phase modulation is effectively equivalent the optimal feedback; this is explained by the fact that the feedback time trace for the two cooling approaches is almost the same (see inset in figure~\ref{figure1}). The difference between the sine profile and the square-wave function does not substantially affect the magnitude of the cooling.

\begin{figure}[t!]
  \centering
  \includegraphics[width=0.45\textwidth]{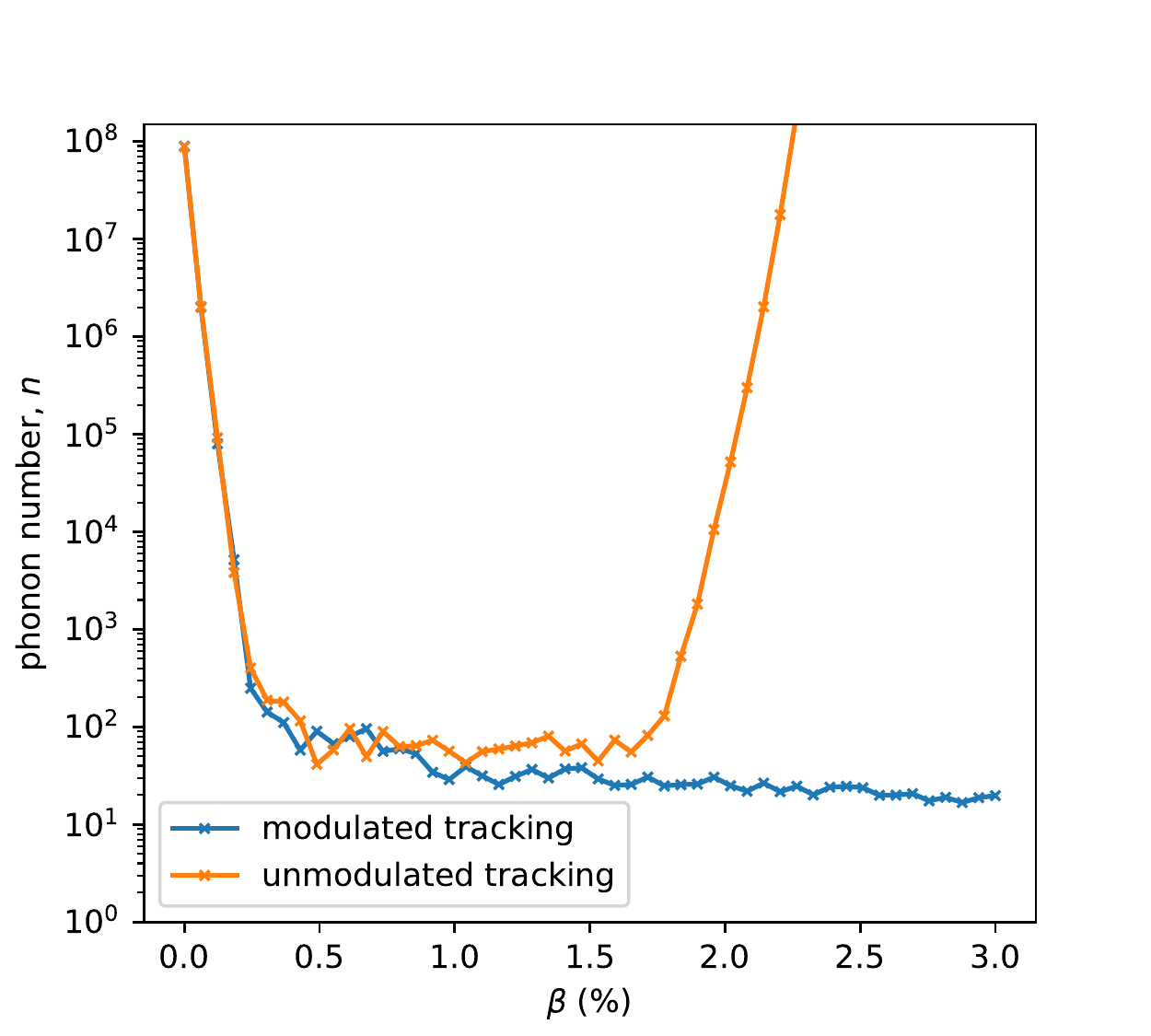}
  \caption{\label{figure2.5}
Quadratic feedback: dependency of the average phonon number reached once the energy has converged with the modulation depth of feedback cooling for double phase quadratic feedback for modulated and unmodulated tracking.}
\end{figure}

However, there is one important difference between the basic tracking technique (exploited in double phase cooling) and the new tracking (exploited in optimal cooling). The first is performed via a Kalman filter that simulates a phase-locked loop (PLL) by exploiting equations~\eqref{x1}-\eqref{x5} with $\beta=0$ (unmodulated tracking). The latter instead makes use of equations~\eqref{x1}-\eqref{x5} with the same $\beta$ as in equations~\eqref{emx1}-\eqref{emx5} (modulated tracking), and requires a fully FPGA-based implementation. One of the limitations of the typical (unmodulated) implementation of parametric feedback is that one can reach only strength of about $\beta=0.01$. This is due to the fact that for higher values of $\beta$ the tracking loses track of the particle because of the larger frequency variation of the system and therefore cooling is not as effective, and for higher modulation depths even heats the system instead of cooling, see Figure \ref{figure2.5}. One of the merits of our improved tracking scheme is that it allows us to track the system trajectories for higher modulation strength. Figures~\ref{figure2} and \ref{figure2.5} clearly show that $\beta$ can be increased, allowing faster cooling rates and lower particle temperatures to be obtained.

\subsection{Linear feedback\label{subsec:Lin feedback}}
Linear feedback can be implemented in the experimental setup by inserting into the vacuum chamber a needle to which an electric voltage is applied~\cite{Timberlake2019}. The electric force generated by the needle affects the particle motion as described in section~\ref{sec:Dynamical models}, i.e. we set $\beta=0$ and modulate the control function $v$. 

\begin{figure}[t!]
  \centering
  \includegraphics[width=0.45\textwidth]{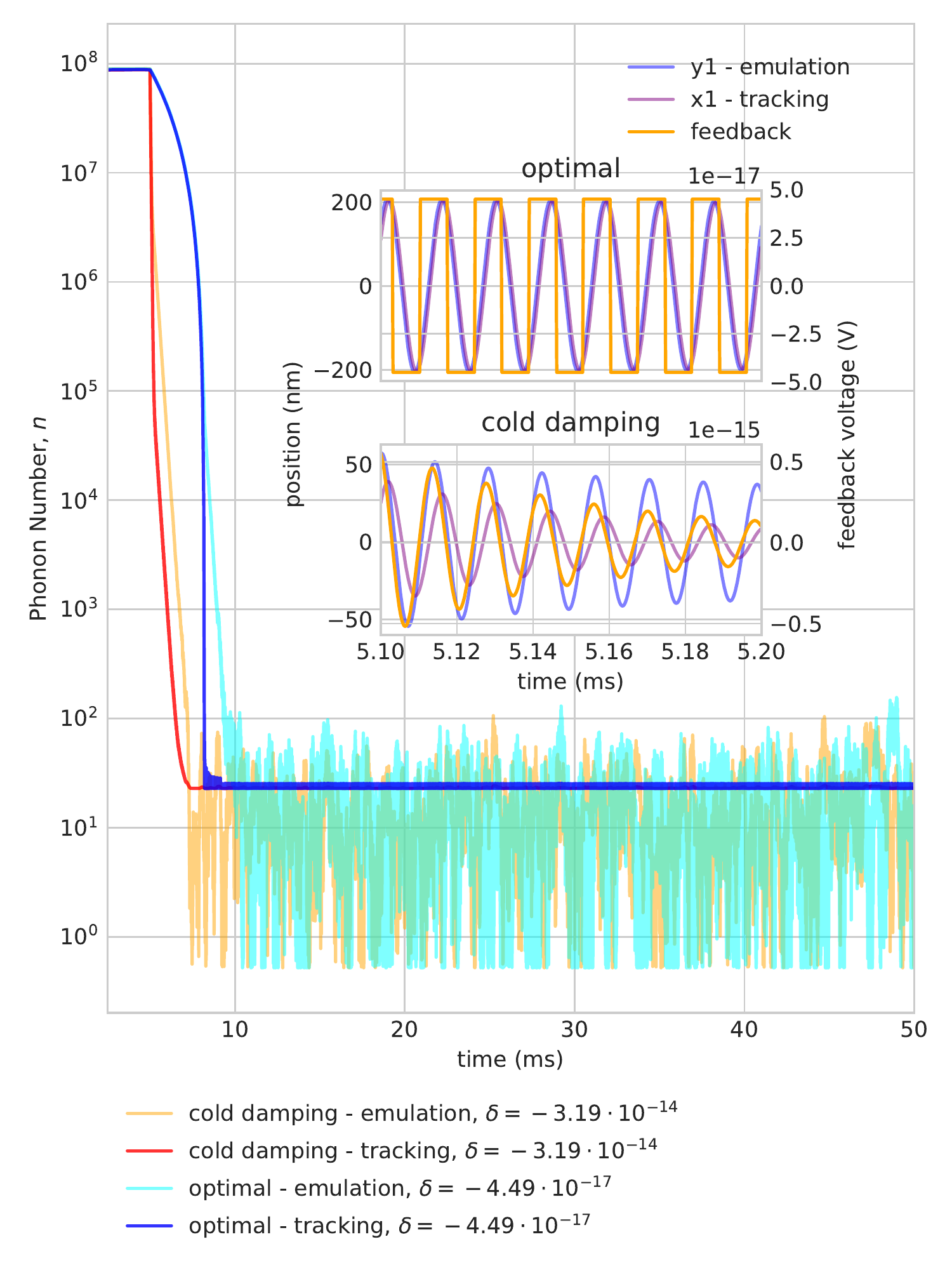}
  \caption{\label{figure3}
    Linear feedback: simulated time trace of the phonon number associated with the translational motion in the emulation and tracking equations. For the first $5\,ms$ the system evolves freely in the harmonic trap before optimal or cold damping linear feedback cooling is applied. The insets show a small slice of the simulated position in the emulation and tracking equations along with the corresponding feedback signal applied to cool the translational motion for optimal and cold damping linear feedback cooling. For these two simulations, different $\delta$ values were chosen such that both are at the minimum in temperature.
}
\end{figure}

Figure~\ref{figure3} compares the optimal linear feedback with cold damping, i.e. a force proportional to the particle velocity, $v=\dfrac{y_2}{m}$ showing that the latter is comparable to the performance of the optimal feedback. The explanation is the same as the quadratic case: the difference between the square optimal signal and the sinusoidal velocity does not significantly affect the cooling efficiency. An important issue one needs to account for while using linear feedback is that the force kicking the particle might lead to re-heating and particle loss if they are too strong. For this reason the simulation of figure~\ref{figure3} makes use of a $\delta$ in the ``optimal range'' identified in figure~\ref{figure3.5}.

\subsection{Discussion\label{subsec:Discussion}}

From our simulation and analysis, it is found that low phonon number states can in principle be reached with both quadratic (figure~\ref{figure1}) and linear (figure~\ref{figure3}) feedback protocols. We find that the cooling signal obtained via optimal control theory does not outperform typical feedback cooling. This can be explained by the fact that our knowledge of the system is given only by the (measured) position of the particle, and this contains all the information about its dynamics (including decoherence and recoil effects). Since typical feedback controls are based on the measured position and velocity of the particle (i.e. the difference of two subsequent positions), they already contain our best knowledge on the system, and the control shape does not play a crucial role. We remark here that the equations for the mean values (equations~\eqref{x1}-\eqref{x2} and \eqref{y1}-\eqref{y2}) are essentially decoupled from the equations for the variances (equations~\eqref{x3}-\eqref{x5} and \eqref{y3}-\eqref{emx5}), the only connection being the arguments of the control functions (see equation~\eqref{controls}). Accordingly, for all practical purposes it is enough to use only equations~\eqref{x1}-\eqref{x2} and \eqref{y1}-\eqref{y2}.

\begin{figure}[t!]
  \centering
  \includegraphics[width=0.45\textwidth]{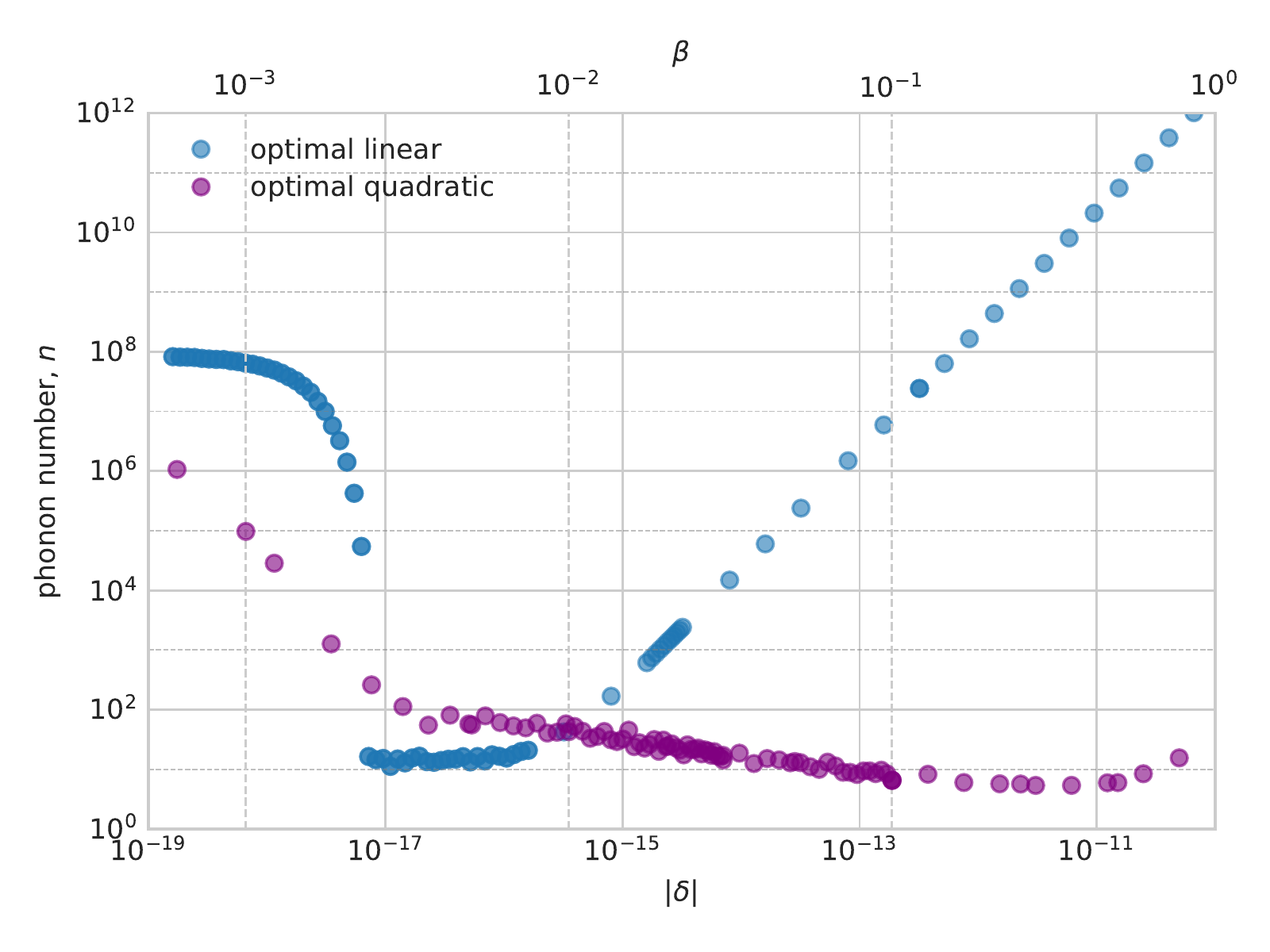}
  \caption{\label{figure3.5}
      Shows the dependency of the phonon number reached on the strength of the feedback applied for optimal linear and optimal quadratic feedback with where the measurement time-step $\Delta t$ is $1\mu s$. For double phase quadratic feedback the data is very similar, for cold damping the graph is also similar, although shifted right by around $10^2$, but temperature diverges to infinity after a critical point, as the cooling strength term in cold damping is proportional to the velocity.}
\end{figure}

We find that for quadratic feedback, increasing the modulation depths $\beta$ decreases the minimum phonon number (or temperature) achievable, and for linear feedback there is an `optimal range' of coupling strength $\delta$ for when cooling is most effective. Above this optimal region the particle will be cooled less effectively, and eventually heated, as $\delta$ increases. The achievable phonon number for increased modulation depth (coupling strength) can be seen in figure~\ref{figure3.5}.


\begin{figure}[t!]
 \centering
  \includegraphics[width=0.45\textwidth]{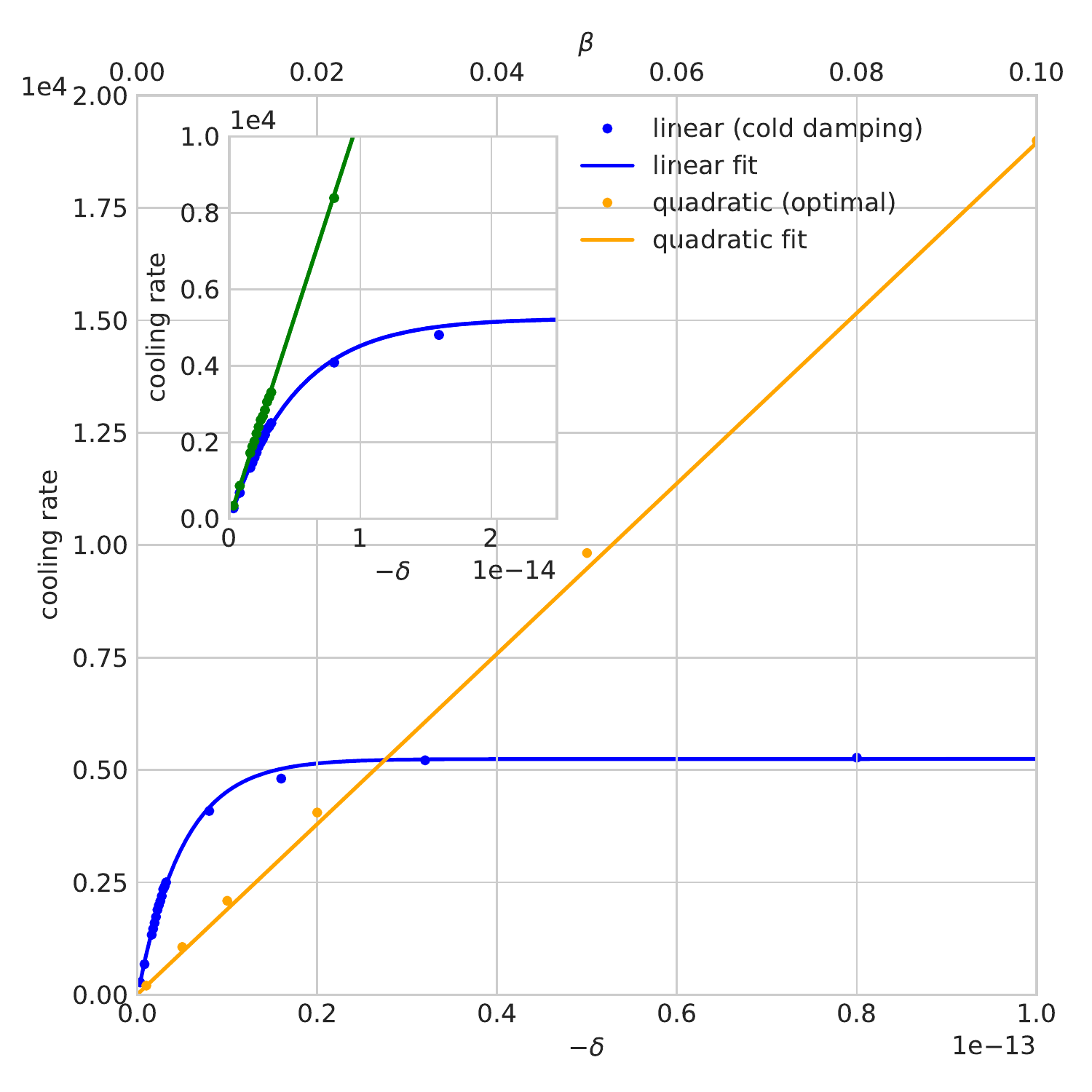}
  \caption{\label{figure3.5.5}
    Cooling rates: Shows the dependency of the initial cooling rate $r_c$ (where the phonon number $n = Ae^{(-r_c t)}$) on the the strength of the feedback applied for linear and quadratic feedback where the measurement time-step $\Delta t$ is $1\mu s$. The inset shows the initial cooling rate with linear cooling where for the blue data the measurement time-step $\Delta t$ is $1\mu s$ and for the green data where the measurement time-step $\Delta t$ is $10ns$.
}
\end{figure}

Increasing the cooling strengths $\beta$ and $\delta$ also increases the initial rate that the nanoparticle is cooled, as expected. The initial cooling rate as a function of cooling strength for both cases can be seen in figure~\ref{figure3.5.5}. As the quadratic modulation depth $\beta$ increases the initial cooling rate increases linearly, whereas increasing the linear cooling strength $\delta$ results in an non-linear increase in the initial cooling rate that approaches an asymptotic value. It was found numerically that this is because the equations~\eqref{emx1}-\eqref{emx5} with $\beta \neq 0, \delta = 0$ (quadratic feedback) require a smaller sampling rate than the equations~\eqref{emx1}-\eqref{emx5} with $\beta = 0, \delta \neq 0$ (linear feedback) in order to track the system well. By increasing the sampling rate sufficiently, it was found that the cooling rate also increases linearly for linear feedback, as is shown in the inset of figure~\ref{figure3.5.5}. Experimentally, where sampling rates cannot be set arbitrarily high, this has practical implications in cooling rates that can be achieved via linear feedback.  


In previous experimental works, where parametric feedback has been utilized, the tracking of the particle's motion does not take into account the laser intensity modulation due to the feedback, which results in a maximum modulation depth of $\sim1.5\%$~\cite{Vovetal}, after which the effectiveness of cooling decreases, eventually causing heating. This is due to the fact that this tracking does not take into account the varying laser intensity, which effectively changes the oscillation frequency of the particle for a fraction of an oscillation period. For small modulation depths this isn't an issue as the effective frequency is still within the tracking bandwidth of the tracking mechanism, but for larger modulations results in the feedback being out of phase. The tracking scheme presented here overcomes this limitation by factoring in the laser modulation in the tracking equations, allowing access to extremely high modulation depths and cooling rates with quadratic feedback. 




After cooling, it was found that the contribution to the mean energy of the particle is dominated by the expectation values of the position and velocity, ($x_1,x_2$ in equation~\eqref{meanE}) whereas the variances' ($x_3,x_4$) contribution is found to be negligible. The energy contained in the variances quite rapidly achieve the Heisenberg limit $x_3x_4=\hbar/2$, and the energy fluctuations are due to random collisions with gas particles and photons. It is interesting to investigate which of these has more of an effect of the particle dynamics. Note that since the variances are constant , equations~\eqref{x3}-\eqref{x5} are essentially a ``time-dependent Kalman filter''~\cite{Kal60,KalBuc61}.

\begin{figure}[t!]
  \centering
  \includegraphics[width=0.45\textwidth]{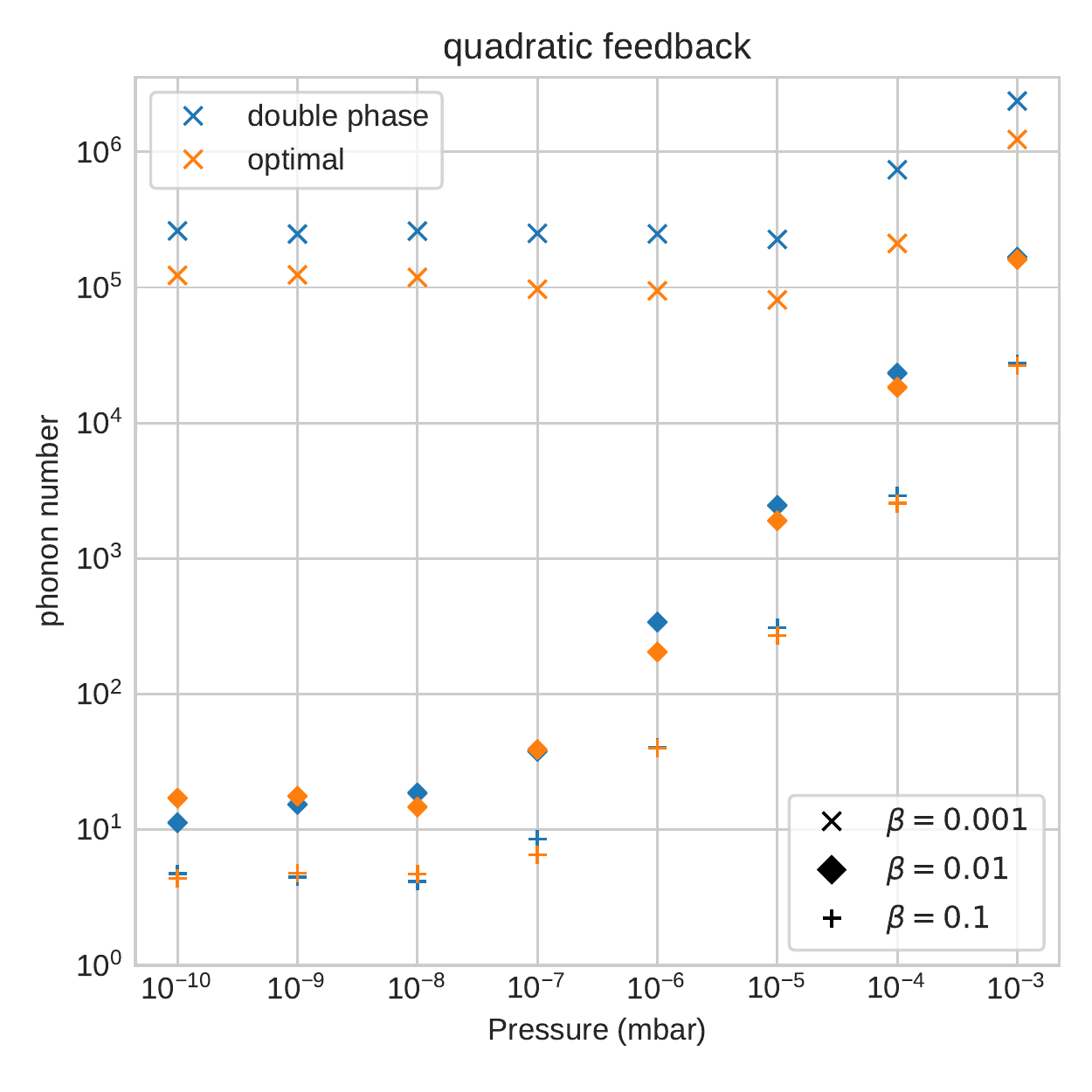}
  \caption{\label{figure5}
    Shows the dependency of the average phonon number reached once the energy has converged with the pressure at which the simulation is performed for optimal and double phase quadratic feedback at three different feedback modulation depths. At about $\sim 10^{-8}$mbar we reach the photon-recoil regime~\cite{Vijay2016}. }
     \end{figure}

Figure~\ref{figure5} shows that for high pressures the phonon number is mostly affected by the gas particles collisions. This effect can be made negligible by reducing the gas pressure in the vacuum chamber. However, when the pressure is sufficiently low one reaches the photon recoil limit, the regime where fluctuations are mostly given by the photon scattering. This kind of effect is always present in the experiment, and it ultimately limits the achievable phonon value. One might try to decrease photon scattering by lowering the laser power, but this leads to a less stable trapping and to a weaker detected signal.


We eventually remark that linear and quadratic feedback can be combined and used at the same time although in our investigations with modulation depth around 1.5\% this has not significantly altered the final temperature. 

\section{Conclusions}
We have considered a cavityless levitation experimental setup, and we numerically investigated both parametric (quadratic) and electric (linear) optimal feedback cooling. The comparison of optimal feedbacks against the typical implementations (respectively double phase and cold damping) show that, although the feedback profiles are different, this does not substantially affect the magnitude of the cooling. However, the implementation of optimal feedback forced us to develop a more sophisticated tracking scheme. This allowed us to go beyond one of the limitations of the typical implementations of parametric feedback, namely the low modulation strength limit. One of the merits of the more sophisticated tracking scheme is that it allows us to increase the modulation strength, obtaining a faster cooling rate and reaching lower temperatures. Furthermore, it was found that for linear feedback there exists an `optimal range' for the coupling strength, that provides most effective cooling.
Moreover, although combined (quadratic+linear) feedback does not seem to significantly alter the achieved temperature, it might still be experimentally helpful to make the cooling more stable. Further improvement might be obtained by applying functional non-Markovian techniques~\cite{FerBas12} to optimal control theory, in order to account for experimental time lags in the derivation of the optimal feedback.

\section*{Acknowledgements}
LF acknowledges funding from the Royal Society under the Newton International Fellowship No NF170345. We thank the EU Horizon 2020 research and innovation program under Grant Agreement No. 766900 [TEQ], the Leverhulme Trust [RPG-2016-046] and the EU COST action CA15220 (QTSpace) for funding support. AS is supported by the United Kingdom Engineering and Physical Sciences Research Council (EPSRC) under Centre for Doctoral Training Grant No. EP/L015382/1.

\end{document}